\documentclass[twocolumn,aps,prl,superscriptaddress,showpacs,secnumroman,showkeys]{revtex4}

\usepackage{CJK}%
\usepackage{color}
\usepackage{amsmath,bm}%
\usepackage{graphicx}%
\begin{document}
%\begin{CJK*}{GBK}{}
\title{Experimental reconstruction of primary hot isotopes and characteristic properties of the fragmenting source in the heavy ion reactions near the Fermi energy}
%\author{W. Lin(ÁÖì¿Æ½)}
\author{W. Lin}
\affiliation{Institute of Modern Physics, Chinese Academy of Sciences, Lanzhou, 730000, China}
\affiliation{University of Chinese Academy of Sciences, Beijing 100049, China}
%\author{X. Liu(ÁõÐÇȪ)}
\author{X. Liu}
\affiliation{Institute of Modern Physics, Chinese Academy of Sciences, Lanzhou, 730000, China}
\affiliation{University of Chinese Academy of Sciences, Beijing 100049, China}
\author{M. R. D. Rodrigues}
\affiliation{Instituto de F\'{\i}sica, Universidade de S\~{a}o Paulo, Caixa Postal 66318, CEP 05389-970, S\~{a}o Paulo, SP, Brazil}
\author{S. Kowalski}
\affiliation{Institute of Physics, Silesia University, Katowice, Poland.}
\author{R. Wada}
\email[E-mail at:]{wada@comp.tamu.edu}
\affiliation{Institute of Modern Physics, Chinese Academy of Sciences, Lanzhou, 730000, China}
%\author{M. Huang(»ÆÃÀÈÝ)}
\author{M. Huang}
%\email[E-mail at:]{huang@comp.tamu.edu}
\affiliation{Institute of Modern Physics, Chinese Academy of Sciences, Lanzhou, 730000, China}
%\author{S. Zhang(ÕÅËÕÑÅÀ­ÍÂ)}
\author{S. Zhang}
\affiliation{Institute of Modern Physics, Chinese Academy of Sciences, Lanzhou, 730000, China}
\affiliation{University of Chinese Academy of Sciences, Beijing 100049, China}
%\author{Z. Chen(³Â־ǿ)}
\author{Z. Chen}
\affiliation{Institute of Modern Physics, Chinese Academy of Sciences, Lanzhou, 730000, China}
%\author{J. Wang(Íõ½¨ËÉ)}
\author{J. Wang}
\affiliation{Institute of Modern Physics, Chinese Academy of Sciences, Lanzhou, 730000, China}
%\author{G. Q. Xiao(Ф¹úÇà)}
\author{G. Q. Xiao}
\affiliation{Institute of Modern Physics, Chinese Academy of Sciences, Lanzhou, 730000, China}
%\author{R. Han(º«Èð)}
\author{R. Han}
\affiliation{Institute of Modern Physics, Chinese Academy of Sciences, Lanzhou, 730000, China}
%\author{Z. Jin(½ùÔöÑ©)}
\author{Z. Jin}
\affiliation{Institute of Modern Physics, Chinese Academy of Sciences, Lanzhou, 730000, China}
\affiliation{University of Chinese Academy of Sciences, Beijing 100049, China}
%\author{J. Liu(Áõ½¨Á¢)}
\author{J. Liu}
\affiliation{Institute of Modern Physics, Chinese Academy of Sciences, Lanzhou, 730000, China}
%\author{P. Ren(ÈÎÅàÅà)}
\author{P. Ren}
\affiliation{Institute of Modern Physics, Chinese Academy of Sciences, Lanzhou, 730000, China}
\affiliation{University of Chinese Academy of Sciences, Beijing 100049, China}
%\author{F. Shi(ʯ¸£¶°)}
\author{F. Shi}
\affiliation{Institute of Modern Physics, Chinese Academy of Sciences, Lanzhou, 730000, China}
\author{T. Keutgen}
\affiliation{FNRS and IPN, Universit\'e Catholique de Louvain, B-1348 Louvain-Neuve, Belgium}
\author{K. Hagel}
\affiliation{Cyclotron Institute, Texas A$\&$M University, College Station, Texas 77843}
\author{M. Barbui}
\affiliation{Cyclotron Institute, Texas A$\&$M University, College Station, Texas 77843}
\author{C. Bottosso}
\affiliation{Cyclotron Institute, Texas A$\&$M University, College Station, Texas 77843}
\author{A. Bonasera}
\affiliation{Cyclotron Institute, Texas A$\&$M University, College Station, Texas 77843}
\affiliation{Laboratori Nazionali del Sud, INFN,via Santa Sofia, 62, 95123 Catania, Italy}
\author{J. B. Natowitz}
\affiliation{Cyclotron Institute, Texas A$\&$M University, College Station, Texas 77843}
%\author{E. J. Kim}
%\affiliation{Cyclotron Institute, Texas A$\&$M University, College Station, Texas 77843}
%\affiliation{Division of Science Education, Chonbuk National University, Jeonju 561-756, Korea}
\author{T. Materna}
\affiliation{Cyclotron Institute, Texas A$\&$M University, College Station, Texas 77843}
%\author{L. Qin(ÇØÀñ¾ý)}
\author{L. Qin}
\affiliation{Cyclotron Institute, Texas A$\&$M University, College Station, Texas 77843}
\author{P. K. Sahu}
\affiliation{Cyclotron Institute, Texas A$\&$M University, College Station, Texas 77843}
%\author{K. J. Schmidt}
%\affiliation{Institute of Physics, Silesia University, Katowice, Poland.}
%\affiliation{Cyclotron Institute, Texas A$\&$M University, College Station, Texas 77843}
%\author{S. Wuenschel}
%\affiliation{Cyclotron Institute, Texas A$\&$M University, College Station, Texas 77843}
%\author{H. Zheng(Ö£»ª)}
\author{H. Zheng}
\affiliation{Cyclotron Institute, Texas A$\&$M University, College Station, Texas 77843}
\affiliation{Physics Department, Texas A$\&$M University, College Station, Texas 77843}
\date{\today}

\begin{abstract}
The characteristic properties of the hot nuclear matter existing at the time of fragment formation in the multifragmentation events produced in the reaction  $^{64}$Zn + $^{112}$Sn at 40 MeV/nucleon are studied.  A kinematical focusing method is employed to determine the multiplicities of evaporated light particles, associated with isotopically identified intermediate mass fragments. From these data the primary isotopic yield distributions are reconstructed using a Monte Carlo method.  The reconstructed yield distributions are in good agreement with the primary isotope distributions obtained from AMD transport model simulations. Utilizing the reconstructed yields, power distribution, characteristic properties of the emitting source are examined.
The primary mass distributions exhibit a power law distribution with the critical exponent, $A^{-2.3}$, for $A \geq 15$ isotopes, but significantly deviates from that for the lighter isotopes. Based on the Modified Fisher Model£¬the ratios of the Coulomb and symmetry energy coefficients relative to the temperature, $a_{c}/T$ and $a_{sym}/T$, are extracted as a function of A. The extracted $a_{sym}/T$ values are compared with results of the AMD simulations using Gogny interactions with different density dependencies of the symmetry energy term. The calculated $a_{sym}/T$ values show a close relation to the symmetry energy at the density at the time of the fragment formation. From this relation the density of the fragmenting source is determined to be $\rho /\rho_{0} = (0.63 \pm 0.03 )$. Using this density, the symmetry energy coefficient and the temperature of fragmenting source are determined in a self-consistent manner as $a_{sym} = (24.7 \pm 3.4) MeV$ and $T=(4.9 \pm 0.2)$ MeV.

\end{abstract}


\maketitle
%\end{CJK*}

\section*{I. Introduction}

In violent heavy ion collisions in the intermediate energy regime (20 $\leq E_{inc} \leq$ a few hundred MeV/nucleon), intermediate mass fragments (IMFs) are copiously produced in multifragmentation processes.  Nuclear multifragmentation was predicted  long ago~\cite{Bohr36} and has been extensively studied following the advent of 4$\pi$ detectors~\cite{Borderie08,Gulminelli06,Chomaz04}. Nuclear multifragmentation occurs when a large amount of energy is deposited in a finite nucleus. The multifragmentation process provides a wealth of information on nuclear dynamics, on the properties of the nuclear equation of state  and on possible nuclear phase transitions. The multifragmentation process was first suggested in the early 1980's~\cite{Finn82,Minich1982,Hirsch1984} as  providing possible evidence for a nuclear matter phase transition\cite{Gulminelli06,Elliott00}. However the specific properties of the nuclear phase transition in hot nuclear matter are still in debate.

The nuclear symmetry energy, a key part of the equation of state, plays an important role in fragment generation in the multifragmentation processas well as in  various phenomena in nuclear astrophysics, nuclear structure, and nuclear reactions. Determination of the density dependence of the symmetry energy is a key objective in many recent laboratory experiments~\cite{Lattimer04,BALi08}. Investigations of the density dependence of the symmetry energy have been conducted using  observables such as isotopic yield ratios~\cite{Tsang2001}, isospin diffusion~\cite{Tsang2004}, neutron-proton emission ratios~\cite{Famiano2006}, giant monopole resonances~\cite{Li2007}, pygmy dipole resonances~\cite{Klimkiewicz2007}, giant dipole resonances ~\cite{Trippa2008}, collective flow~\cite{zak2012} and isoscaling~\cite{Xu2000,Tsang2001_1,Huang2010}. Different observables may probe the properties of the symmetry energy at different densities and temperatures.

In general, the nuclear multifragmentation process can be divided into three stages, i.e., dynamical compression and expansion, the formation of primary hot fragments, and finally the separation and cooling of the primary hot fragments by evaporation.
%\textcolor{blue}{
To model the multifragmentation process, a number of different models have been developed since Boltzmann-Uehling-Uhlenbeck(BUU) model~\cite{Aichelin85}, a test particle based Monte Carlo transport model, was coded in 1980's. Stochastic mean field (SMF)~\cite{Colonna98,Baran12,Gagnon12}, Vlasov-Uehling-Uhlenbeck model(VUU)~\cite{Kruse85}, Boltzmann-Nordheim-Vlasov model(BNV) 	~\cite{Baran02} are also based on the test particle method. Instead of using the test particles, Gaussian wave packets are introduced in quantum molecular dynamics such as quantum molecular dynamics model (QMD)~\cite{Peilert89,Aichelin91,Lukasik93}. Constrained molecular dynamics(CoMD)~\cite{Papa01,Papa05,Papa07,Papa09} and improved quantum molecular dynamics model (ImQMD)~\cite{Wang02,Wang04,Zhang05,Zhang06,Zhang12} are based on QMD, but an improved treatment is made on the Pauli blocking during the time evolution of the reaction. Fermionic molecular dynamics(FMD)~\cite{Feldmeier90} and antisymmetrized molecular dynamics (AMD)~\cite{Ono96,Ono99,Ono02} are most sophisticated models, in which the Pauli principle is taken into account in an exact manner in the time evolution of the wave packet and stochastic nucleon-nucleon collisions.
%}
Most of them can account reasonably well for many characteristic properties experimentally observed. On the other hand statistical multifragmentation models such as microcanonical Metropolitan Monte Carlo model (MMMC)~\cite{Gross90,DAgostino99} and statistical multifragmentation model(SMM)~\cite{DAgostino99,Bondorf85,Bondorf95,Botvina95,DAgostino96,Scharenberg01,Bellaize02,Avdeyev02,Ogul12},  based on a quite different assumption from the transport models, can also describe many experimental observables well. The statistical models use a freeze-out concept. The multifragmentation is assumed to take place in equilibrated nuclear matter described by parameters, such as size, neutron/proton ratio, density and temperature. In recent analyses the parameters are optimized to reproduce the experimental observables of the final state. In contrast, the transport models do not assume any chemical or thermal equilibration. Nucleons travel in a mean field experiencing nucleon-nucleon collisions subject to the Pauli principle. The mean field parameters and the in-medium nucleon-nucleon cross sections are the main physical ingredients. Fragmentation mechanisms also differ from those of the statistical models.

One of the complications one has to face in comparing the model predictions to the experimental observables in either dynamical or statistical multifragmentation models, is the secondary decay process. Multifragmentation is a very fast process which occurs in times of the  order of 100 fm/c, whereas the secondary decay process is a very slow process. When fragments are formed in the multifragmentation process, many may be in excited states and will subsequently cool by secondary decay processes before they are detected~\cite{Marie98,Hudan03,Rodrigues13,Lin14}. The secondary cooling process may significantly alter the fragment yield distributions. Even though the statistical decay process itself is rather well understood and well coded, it is not a trivial task to combine it with a dynamical code. The statistical evaporation codes assume  nuclei at thermal equilibrium with normal nuclear densities and shapes. These conditions are not guaranteed for fragments when they are formed in the multifragmentation process of the primary. We will call the fragments at the time of formation "primary" fragments.  Those observed after the cooling process will be called the observed or "final" fragments~\cite{Huang10_1,Huang10_2,Chen10}.

In order to avoid the complications introduced by the secondary decay and make the comparisons between the experimental data and the results from different models more straight forward, we proposed a kinematical reconstruction of the primary fragment yields. In the previous work of Ref.~\cite{Rodrigues13}, we focused on the kinematical focusing method and the reconstruction of the excitation energy of the primary fragments. In this article the characteristic properties of the fragmenting source are further investigated. A model study of the self-consistent method we have used in determination of the properties of fragmenting source has been described in a letter form in Ref.~\cite{Lin14}.
This article is organized as follows: The experimental procedure is described in Sec. II. The data analysis and the reconstruction of the multiplicity of the primary hot fragments are given in Sec. III. Utilising the reconstructed isotope yields, power low distribution is discussed in Sec. IV. Characteristic properties of the fragmenting system is studied in Sec. V. A brief summary is made in Sec. VI.

\section*{II. Experiment}

The experiment was performed at the K-500 superconducting cyclotron facility at Texas $A\&M$ University. $^{64,70}$Zn and $^{64}$Ni beams were used to irradiate $^{58,64}$Ni, $^{112,124}$Sn, $^{197}$Au, and $^{232}$Th targets at 40 MeV/nucleon. In this article, we focus on the $^{64}$Zn + $^{112}$Sn reaction, which had the best statistical precision. Details of the experiment have been given in Refs.~\cite{Huang10_2,Rodrigues13,Zhang13}. Here we briefly outline the experiment and clarify some issues. Intermediate mass fragments (IMFs; 3 $\leq$ Z $\leq$ 18) were detected by a detector telescope placed at $\theta_{lab} = 20^{o}$.
The telescope consisted of four Si detectors. Each Si detector had an effective area of 5 x 5 $cm^2$. The nominal detector thicknesses were $129, 300, 1000,$ and $1000 \mu m$. All Si detectors were segmented into four sections and each quadrant subtended $5^{o}$ in the polar angle. Typically six to eight isotopes for 3 $\leq$ Z $\leq$ 18 were clearly identified using the $\Delta E$ x E technique employing any two consecutive detectors. Mass identification of the isotopes was verified using a range-energy table~\cite{Hubert90}. The laboratory energy thresholds ranged from  4 to 10 MeV/nucleon, from Li isotopes to the heaviest isotopes identified.

Two sets of detectors were used to detect the light particles. For the light charged particles (LCPs), 16 single-crystal CsI(Tl) detectors of 3 cm length were set around the target at angles between $\theta_{Lab} = 27^{o}$ and $155^{o}$, tilted $30^{o}$ in the azimuthal angle to avoid shadowing the neutron detectors described below. The light output from each detector was read by a photomultiplier tube. The pulse shape discrimination method was used to identify p, d, t , $^{3}He$ and $\alpha$ particles. The energy calibrations for these particles were performed using Si detectors of $50$ to $300 \mu m$ in front of the CsI detectors in a separate run.

\begin{figure}
\includegraphics[scale=0.4]{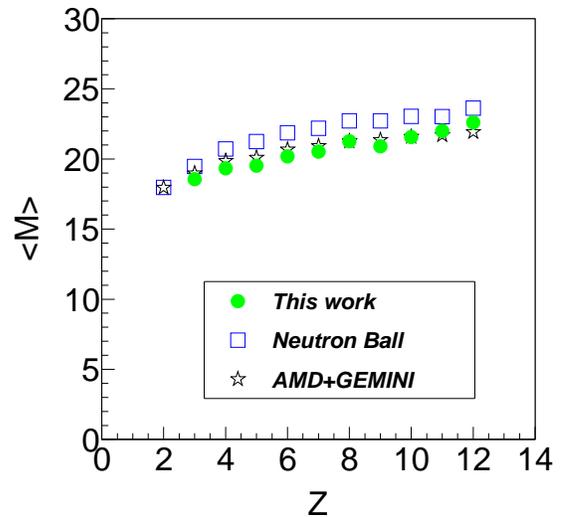}
\caption{\footnotesize
(Color online)Comparisons of total neutron multiplicity obtained from this work (dots), neutron ball measurement (squares) and from an AMD+GEMINI calculation (stars). The figure is taken from Ref.~\cite{Rodrigues13}.
 }			
\label{fig:fig5_Mtot_neutrons}
\end{figure}

For neutrons 16 detectors of the Belgian-French neutron detector array, DEMON, were used~\cite{Tilquin95}. The  set up of the neutron detectors is described in detail in Ref.~\cite{Zhang13}. Eight of them were set in the plane perpendicular to the reaction plane. The zero degree in polar and azimuthal angles of the opening angle was taken to be the telescope direction.
%\textcolor{blue}{
The reaction plane of the neutron distribution from the observed IMF is defined by the vector of the telescope direction and that of the beam.
%}
The other eight neutron detectors were set in the reaction plane. The detectors were distributed to achieve opening angles between the telescope and the DEMON detector of $15^{o} \le \theta_{IMF-n} \le 160^{o}$. Neutron/gamma discrimination was obtained from a pulse shape analysis, by comparing the slow component of the light output to the total light output. The neutron detection efficiency of the DEMON detector, averaged over the whole volume, was calculated using GEANT and applied to determine neutron multiplicities~\cite{Zhang13}. The derived multiplicities from this experiment are shown in Fig.\ref{fig:fig5_Mtot_neutrons}, taken from Ref.~\cite{Rodrigues13}. In that figure they are also compared to results obtained in a separate experiment for the same reaction using the neutron ball calorimeter in the NIMROD detector array and to results of an AMD+GEMINI simulation of this reaction~\cite{Ono99,Charity88}.

\begin{figure}
\includegraphics[scale=0.4]{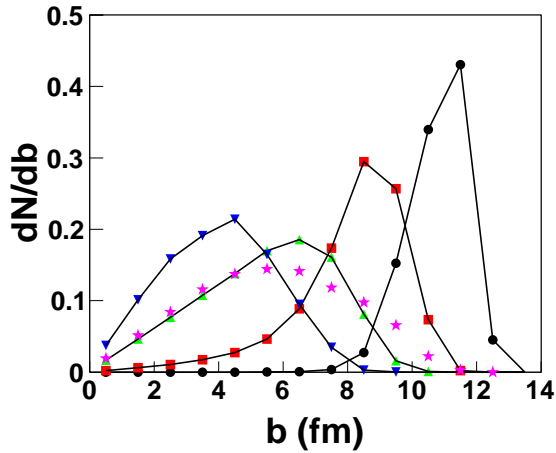}
\caption{\footnotesize
(Color online) Simulated impact parameter distributions for violent (downward  triangles), semi-violent (upward triangles), semi-peripheral (squares) and peripheral (dots) collisions. Stars indicate the events in which at least one IMF $(Z \ge 3)$ is emitted at 20$^\circ \pm 5^\circ$. The summed distribution for a given class is normalized to 1. The figure is taken from Ref.~\cite{Huang10_2}.
}			
\label{fig:fig_Bimp}
\end{figure}

In the experiment, the telescope at $\theta_{lab} = 20^{o}$ was used as the main trigger. The angle of the telescope was optimized to be small enough so that sufficient IMF yields are obtained above the detector energy threshold, but large enough so that the contribution from peripheral collisions was negligible according to AMD+GEMINI simulations.
%\textcolor{blue}{
The events triggered by IMFs in this experiment are "inclusive", but they belong to a certain class of events. In order to determine the event class taken in this experiment, AMD simulations are used
to evaluate the impact parameter range sampled and the IMF production mechanism involved in the present data set.
In Fig.\ref{fig:fig_Bimp}, calculated impact parameter distributions are presented. The violence of the reaction for each event in the AMD simulation is determined in the same way as our previous work~\cite{Wada04}, in which the multiplicity of light particles, including neutrons, and the transverse energy of light charged particles were used.
%}
The resultant impact parameter distributions are shown for each class of events together with that of the events in which at least one IMF is emitted at angle of 20$^\circ \pm 5^\circ$. As seen in the figure the distribution of the events selected by the IMF detection is very similar to that for semi-violent collisions which have a broad impact parameter distribution overlapping significantly with that of violent collisions.
%\textcolor{blue}{
The event class identification in this experiment is crucial for the following analysis. As shown in Ref.~\cite{Wada04}, the IMFs from the semi-violent collisions are dominated in the intermediate velocity(IV) component in the moving source analysis discussed below. Therefore in the following analysis it is assumed that the majority of the events triggered by IMFs at $20^{o}$ in this experiment are representative of the IV source component in the semi-violent collisions.
%}

\begin{figure}
\includegraphics[scale=0.4]{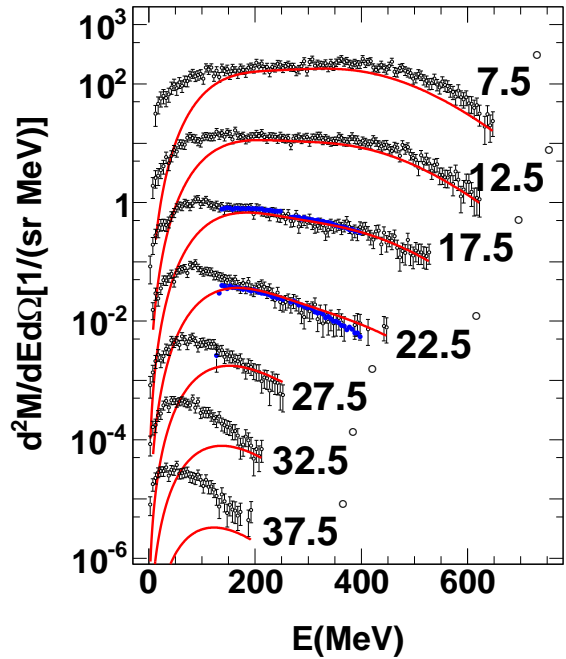}
\caption{\footnotesize
(Color online)Experimental $^{16}O$ energy spectra (closed circles) are compared with the AMD+Gemini result (open circles) for $^{64}Zn + ^{112}$Sn at 40 MeV/nucleon. The spectra for the AMD+Gemini result is obtained for the semi-violent collisions. Detection angles are given in the figure and the absolute Y scale is correspond to the bottom spectra and the spectra are multiplied by a factor of 10 from the bottom to the top. The curves are the result of the moving source fit, in which the parameters are determined from the experimental spectra at 17.5$^\circ$ and 22.5 $^\circ$. The source velocity of $V_s=0.62V_{p}$ and $\Delta V_s = 0.11V_{p}$ are used. The figure is taken from Ref.~\cite{Huang10_2}.
}			
\label{fig:fig_E_Exp_AMD_MS}
\end{figure}

Based on the assumption above, a moving source fit was employed to fit the observed spectra~\cite{Awes81}. For the light particles, three sources, the projectile-like (PLF), the intermediate velocity (IV) and the target-like (TLF)  sources, were assumed. For IMFs, a single IV source was used to extract the multiplicity. In Fig.\ref{fig:fig_E_Exp_AMD_MS}, the experimental energy spectra of $^{16}O$ are compared with the results from an AMD+Gemini calculation in an absolute scale, together with the moving source fit result. The spectra for the AMD+Gemini result are those corresponding to the semi-violent collisions. The experimental spectra at 17.5$^\circ$ and 22.5$^\circ$ are reproduced reasonably by the AMD+Gemini simulation. The moving source parameters were determined from the experimental spectra. For IMFs, a fixed apparent temperature of 17 MeV was used. The IV source velocity was smeared between $V_s \pm \Delta V_s$. Typically $V_s = 0.6V_{p}$ and $\Delta V_s =0.1V_{p}$ were used, where $V_{p}$ is the projectile velocity, but for each case these values were optimized. The majority of the spectra at angles, $\theta \le 20^{\circ}$, are well reproduced by the IV source component, except for the lower energy side of these spectra and those at $\theta \ge 25^{\circ}$. These are attributed to the TLF component. One can also see a small enhancement in the AMD+Gemini result above the moving source fit at forward angles, which is attributed to the PLF source component. For the semi-peripheral or peripheral collisions, a prominent PLF component with the source velocity, $Vs \sim 0.9V_{p}$ appears at forward angles. These are generally observed for all isotopes measured in the reaction presented here. In the following analysis, only the IV source component is taken into account.
%The PLF and TLF source components are not taken into account in the present analysis. Therefore the IMF multiplicities presented in this work are those of the IV source component.

\begin{figure}
\includegraphics[scale=0.4]{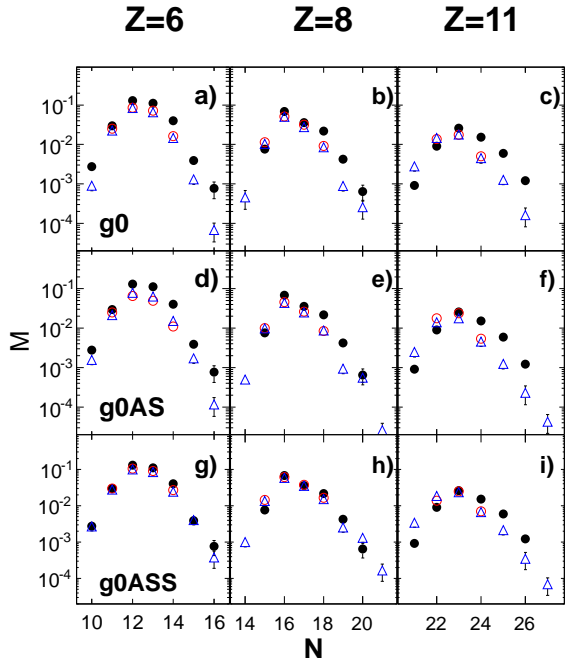}
\caption{\footnotesize
(Color online)Typical cold isotope distributions are compared with the results of AMD+GEMINI simulations with three different interactions discussed in this paper. The results for Z=6, 8 and 11 are plotted from left to right column. From the top to bottom row, the results of AMD with g0, g0AS and g0ASS are plotted with the experimental data, respectively. The experimental data is taken from the IV source component from the moving source fit and shown by dots. Same experimental data are used in each column. The multiplicity distributions from the AMD simulations are calculated in two ways. The circles represent the results of the IV source component from the moving source fit. The triangles are those calculated from the approximated method (see details in the text).
}			
\label{fig:Secondary}
\end{figure}

%\textcolor{blue}{
In Fig.~\ref{fig:Secondary}, the typical experimental cold isotope distributions are compared with those of the AMD simulations. The experimental data are the IV source component from the moving source fit, described above. For the AMD simulations, the IV multiplicities are calculated in two ways, one from the moving source fit and the other by an approximated method. The approximated method is used because of the poor statistics in the yields for the neutron-rich or proton-rich isotopes. As seeing the moving source fit in Fig.~\ref{fig:fig_E_Exp_AMD_MS}, the IV source component dominates in the energy range of $E/A > 5 MeV/nucleon$ and in the angular range of $\theta < 25^{o}$. The TLF component dominates in the energy range of $E/A \le 5 MeV/nucleon$ in the entire angular range shown in the figure. The PLF component is barely seen only in the high energy range at $\theta = 5^{o}$. The PLF contribution becomes significant at $\theta < 5^{o}$ for isotopes with A > 25. Therefore in the approximated method the IV component is calculated by integrating yields at $E/A > 5 MeV/nuceon$ and $5^{o} < \theta < 25^{o}$. Same energy and angular ranges are used for all isotopes. The calculated IV multiplicities in this method are compared with those of the moving source fit in Fig.~\ref{fig:Secondary} for all AMD simulations. Good agreements are obtained for all cases in which the moving source results are available.
%}

\section*{III. Reconstruction of the primary hot isotopes}

Yields of primary hot isotopes have been reconstructed, employing a kinematical focusing technique.
%\textcolor{blue}{
In Fermi energy heavy ion collisions, light particles are emitted at different stages of the reaction and from different sources during the evolution of the collisions. Those from an excited isotope are kinematically focussed into a cone centered along the isotope direction. The kinematical focussing technique uses this nature.
The particles emitted from the precursor fragment of a detected isotope will be called "correlated" particles and those not emitted from the precursor fragment are designated as "uncorrelated" particles. To reconstruct the yield distributions of the primary hot isotopes, it is crucial to distinguish the correlated particles from the uncorrelated particles. When particles are emitted from a moving parent of an isotope (whose velocity is approximated by the velocity of the trigger IMF, $v_{IMF}$), the isotropically emitted particles tend to be kinematically focused into a cone centered along the $v_{IMF}$ vector. In the actual analysis, moving source fits are employed to isolate the correlated light particles, including neutrons, from the uncorrelated ones and the correlated light particle multiplicities are extracted for each isotopes identified in the telescope.
%}
The shape of the uncorrelated spectrum is obtained from the particle velocity spectrum observed in coincidence with Li isotopes, which is the minimum Z of the particle identified in the triggering telescope and associated with least particle emissions\cite{Marie98,Hudan03}. Since the Li associated spectrum includes some pre-cursor decay, the multiplicity extracted for a given isotope needs to be corrected by addition of an amount corresponding to the correlated particle emission from the Li isotopes. This correction has been made using results from the AMD+GEMINI simulation. The amount of the correction was determined by averaging over values obtained in calculations using  different EOS (Gogny interaction of hard and soft EOS) and different versions of the code (AMD/D~\cite{Ono99} and AMD/DS~\cite{Ono02}). Most of the extracted values agreed with each other within a rather small margin.
These values are $0.40\pm0.05$ for neutrons, $0.24\pm0.04$ for protons, $0.044\pm0.005$ for deuterons, $0.035\pm0.005$ for tritons and $0.32\pm0.04$ for $\alpha$ particles. The errors are evaluated from the standard deviations for the different calculations.
The multiplicity of $^{3}He$ was not extracted in this experiment, because of the poor statistics reflecting the much smaller multiplicities than those of the deuterons and tritons.  Therefore $^{3}He$ was not taken into account in the reconstruction analysis.
%\textcolor{blue}{
A further detailed description of the kinematical focusing analysis is given in Ref.~\cite{Rodrigues13}.
%}

\begin{figure}[htb]
\includegraphics[scale=0.4]{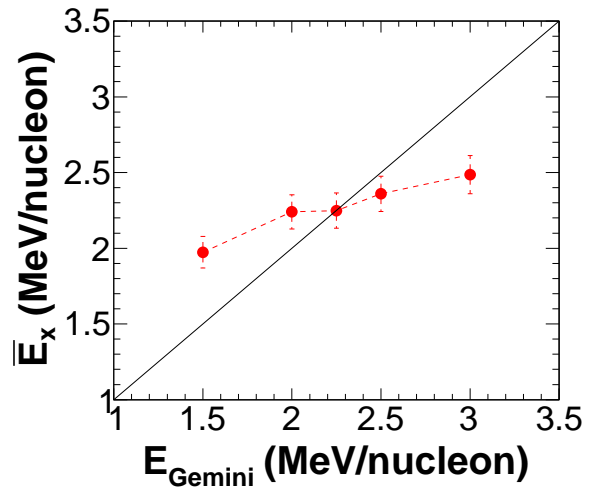}
\caption{\footnotesize
(Color online) The average reconstructed excitation energy vs GEMINI input excitation energy.
}			
\label{fig:figEx}
\end{figure}

The excitation energy and multiplicity distributions of the primary hot isotopes were reconstructed using a Monte Carlo method,
%\textcolor{blue}{
assuming the light particle emissions from an excited isotope are independent each other.
%}
Since only the average values of LP multiplicities can be extracted from this experiment, the shape (centroid and width) of the multiplicity distributions, assuming Gaussian distributions, have been taken from results of the statistical decay code, GEMINI~\cite{Charity88}. The shape depends on the input excitation energy values of the GEMINI calculation. Several input values were used to reconstruct the excitation energy~\cite{Rodrigues13}.
%\textcolor{blue}{
In Fig.\ref{fig:figEx} the average excitation energy per nucleon was calculated for isotopes with $Z \ge 6$, using their multiplicities as weighting factors. The resultant average excitation energies are compared with the input value of the GEMINI calculations and plotted. The input value and the extracted average energy coincide at $E_{x} \sim2.25$ MeV/nucleon and therefore in the following analysis, the input value of $E_x =2.25 \pm 0.25 $ MeV/nucleon was used for the GEMINI calculations to determine the shape of the multiplicity distribution.
%}

\begin{figure}[htb]
\includegraphics[scale=0.4]{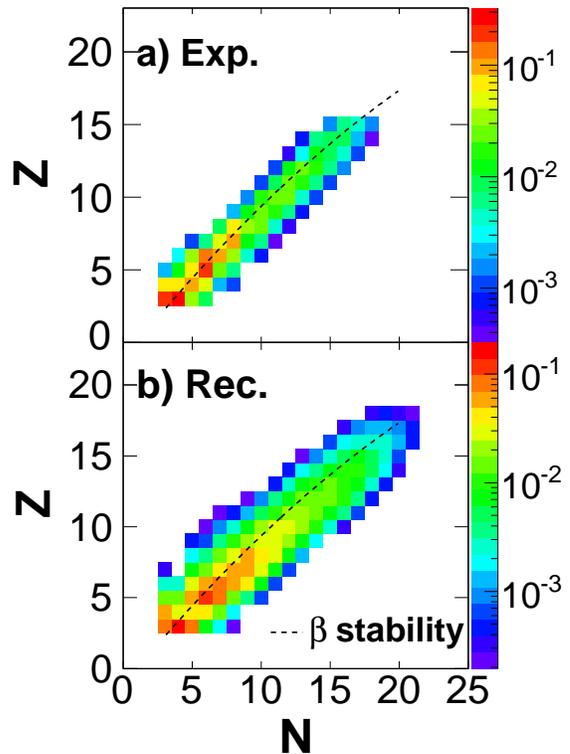}
\caption{\footnotesize
(Color online) Isotope distribution on a 2D plot of Z vs N a) for the experimental and b) the reconstructed fragments. Dashed line indicates the $\beta$ stability line. The Z axis is the multiplicity given in an absolute logarithmic scale.
}			
\label{fig:NZ_distribution}
\end{figure}

The LP multiplicity distributions, $M_{i}(i=n,p,d,t,\alpha)$,  associated with a given detected daughter nucleus were generated  on an event by event basis. For a given width of the Gaussian distribution, generated by the GEMINI simulation, their centroid is adjusted to give the same average multiplicity as that of the experiment.
Using these LP multiplicities, the mass and charge of the primary hot isotopes with $A_{hot}$ and $Z_{hot}$ are calculated as,
\begin{equation}
\begin{split}
A_{hot}=\sum_{i} M_{i} A_{i} + A_{cold} \\
Z_{hot}=\sum_{i} M_{i} Z_{i} + Z_{cold}
\end{split}
\label{eq:mass_charge_hot}
\end{equation}
$A_{i}, Z_{i}$ are the mass and charge of correlated particle i and $A_{cold}, Z_{cold}$ are those of the detected cold isotope.
%Using the Z and A of the experimentally observed cold IMF and the LP event by event multiplicities, the distribution of parent nuclei is reconstructed.
100,000 parents are generated for each experimentally observed isotopes and added with the experimental multiplicity as a weighting factor. The multiplicities associated with the unstable nuclei of $^{8}Be$ and $^{9}B$ were added artificially by estimating their multiplicity and associated LP multiplicities from the neighboring isotopes. In Fig.\ref{fig:NZ_distribution}, the isotopic distributions of the experimentally observed fragments and of the reconstructed hot fragments are shown in 2D plots of Z vs N. The reconstructed primary distributions are significantly broader than those of the experimental cold fragments.

\begin{figure*}[htb]
\includegraphics[scale=0.7]{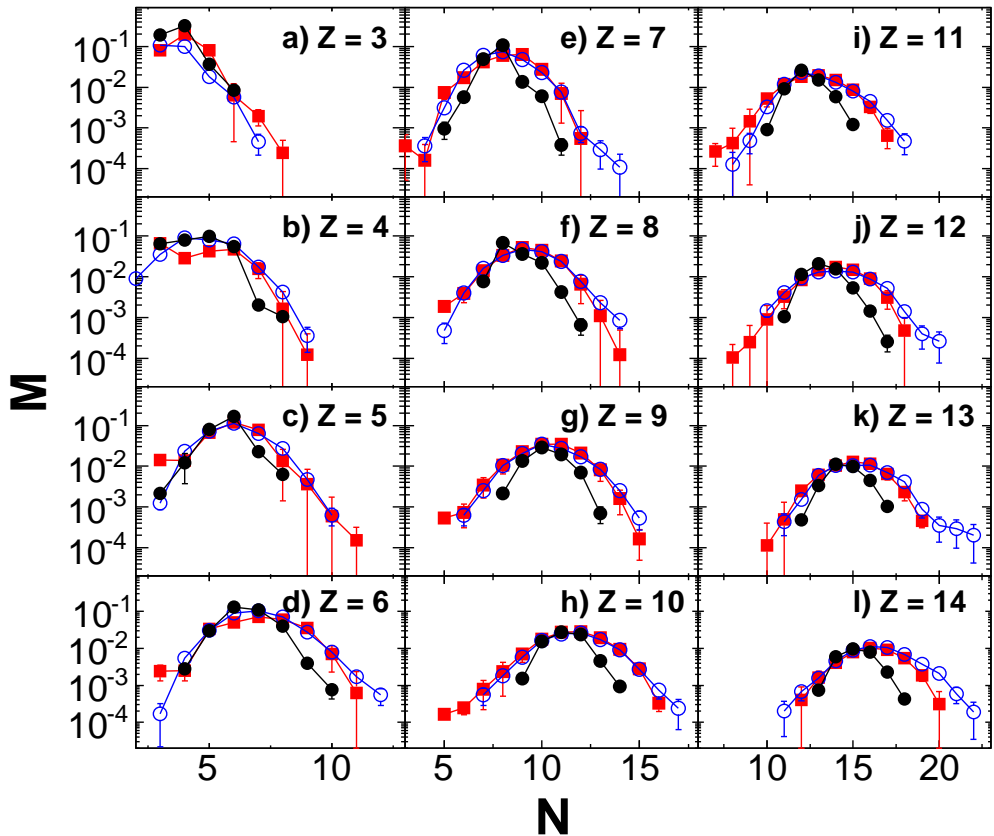}
\caption{\footnotesize
(Color online) Isotopic multiplicity distributions of experimental cold fragments (closed circles), reconstructed hot fragments (closed squares) as well as AMD primary hot fragments as a function of fragments mass number $A$ for a given charge Z, which is indicated in the figure. AMD results are from those of the g0AS interaction.
}			
\label{fig:Isotopic_distribution}
\end{figure*}

\begin{figure}[htb]
\includegraphics[scale=0.35]{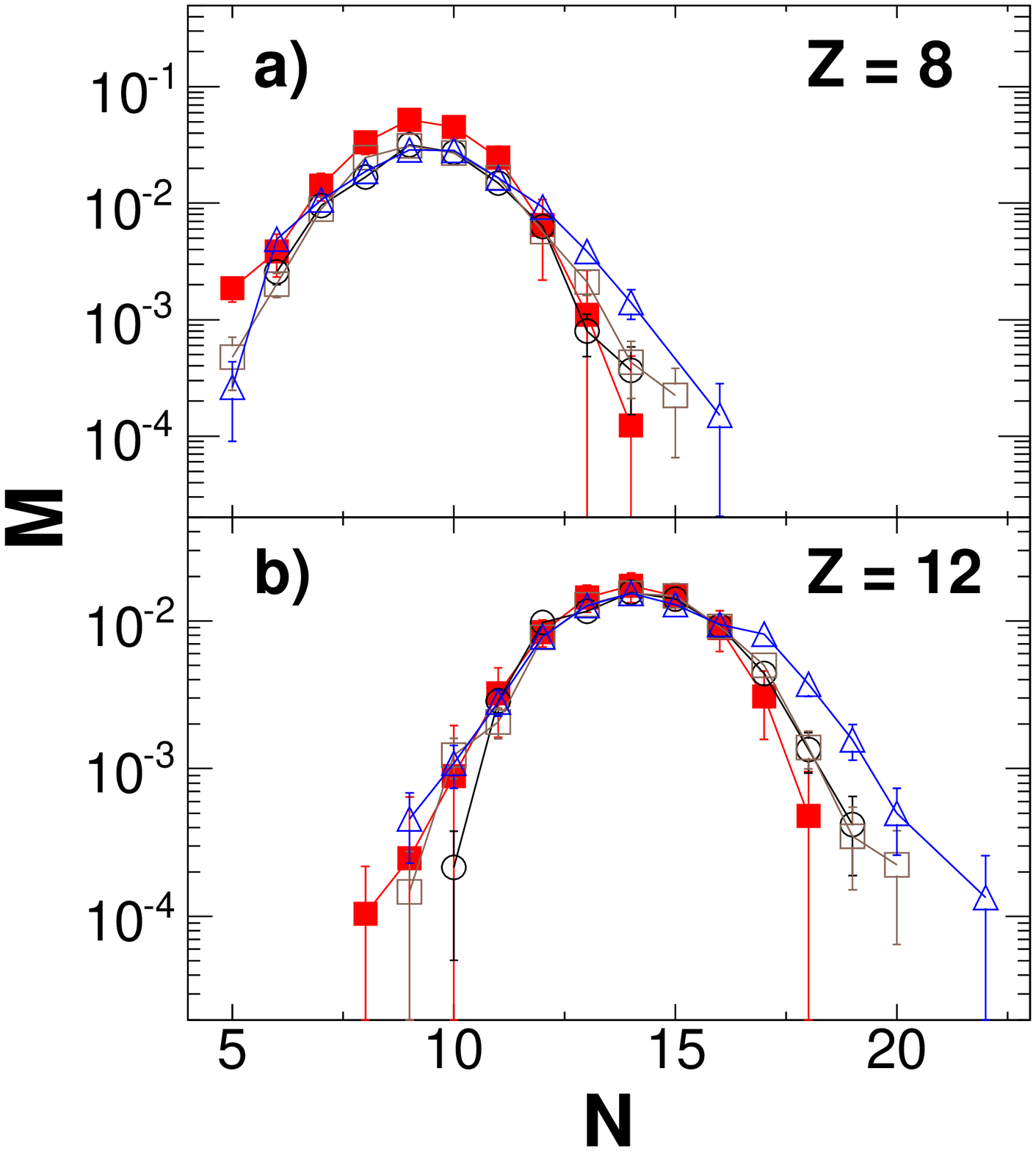}
\caption{\footnotesize
(Color online) Isotopic multiplicity distributions of a) Z=8 and b) Z=12 for the reconstructed hot fragments (closed squares) as well as AMD primary hot fragments with g0 (open circles), g0AS (open squares) and g0ASS (open triangles) as a function of fragment mass number $A$ for a given charge Z, which is indicated in the figure.
}			
\label{fig:Isotopic_distribution_1}
\end{figure}

In Fig.\ref{fig:Isotopic_distribution} the multiplicity distributions of the reconstructed hot isotopes for each charge Z are shown together with the experimentally observed distributions. These are compared to the multiplicity distributions for the AMD primary fragments evaluated at $t = 300 fm/c$. At that time the clusters were identified using a standard coalescence technique with a coalescence radius in phase space of $R_c=5$.

%Since the experimental data are taken from the intermediate velocity (IV) source component of the semi-violent collisions. In order to do a similar source fit, one needs a lot of statistics. For the simulated cold fragments, one AMD event is used 100 times in the GEMINI calculation in order to sample all possible decay channels and therefore one has enough statistics for the moving source analysis. On the other hand for the primary fragments, it is too time consuming to get enough statistics for the source fit at the present CPU power available. Therefore for AMD primary hot fragments, an approximate method is employed, in which the multiplicity of the IV source component is calculated by integrating the energy spectra over $E > 5$ MeV/nucleon and between $5^{\circ} \leq \theta \leq 25^{\circ}$ in the laboratory frame in order to minimize the contribution from the projectile-like and the target-like sources, based on the moving source analysis in Fig.\ref{fig:fig_E_Exp_AMD_MS}. The values of the energy threshold and the angular range for the integration are determined to reproduce the IV source multiplicity of the moving source fit for the energy spectra of the secondary cold isotopes. In this way one can evaluate the multiplicity of the intermediate source to an accuracy of about 5$\%$. For the cold fragment yields from the AMD+GEMINI results, the same moving source analysis is performed and the IV source component is used for the comparison to the experimental multiplicity.
%\textcolor{blue}{
In order to determine the IV source multiplicity for the AMD primary isotopes, the approximated method described in Se.II. is employed, assuming the energy and angular distributions of the primary isotopes are similar to those of the secondary cold isotopes. This assumptions are reasonable because the secondary emissions are isotropic in the GEMINI simulation.
%}
For the selection of semi-violent collisions, the events in the impact parameter range of 0 - 8 $fm$ are used. More than $75\%$ of events in this range belong  to the semi-violent collisions as seen in Fig.\ref{fig:fig_Bimp}.

In Fig.\ref{fig:Isotopic_distribution} comparisons are made in absolute multiplicity.
The reconstructed yields (closed squares) show much wider distributions than those of the cold isotopes (dashed lines), which reflects the significant modification of the primary hot yield caused by the secondary decay process.
The reconstructed yield distributions are compared with the yields of  primary fragments from the AMD simulations.
Overall, the reconstructed primary isotope distributions are reasonably well reproduced by the AMD simulations.
In Fig.\ref{fig:Isotopic_distribution_1} the reconstructed isotope distributions for Z=8 and Z=12 are further compared with primary distributions calculated with the standard Gogny interactions, i.e.,  g0, which has an asymptotic soft symmetry energy,  g0AS  with an asymptotic stiff symmetry energy and g0ASS with an asymptotic super-stiff symmetry energy ~\cite{Ono03}.
%Different choices of the density dependence of the symmetry energy term give notable differences for the very neutron or proton rich isotopes.

\begin{figure}[htb]
\includegraphics[scale=0.35]{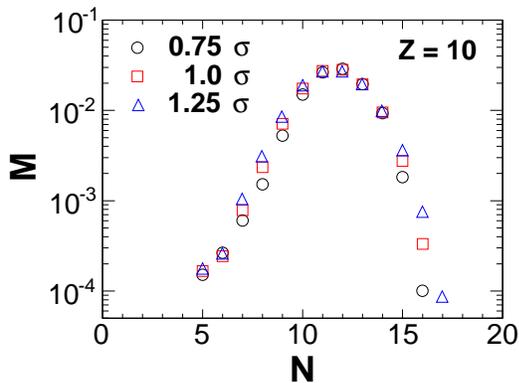}
\caption{\footnotesize
(Color online) Calculated Ne isotope distributions when the $\sigma$ values of the LP multiplicity distribution are changed from for $0.75\sigma$ to $1.25\sigma$.
}			
\label{fig:Sigma}
\end{figure}

The errors of the reconstructed multiplicities in Fig.\ref{fig:Isotopic_distribution} consist of the errors on the associated LP multiplicities from the moving source fit and the errors in the amount added for the correction for the emission from the Li isotopes. Most of the combined errors are at most ~10-20\%. For some of very neutron or proton rich isotopes, a larger contribution of the additional error in the reconstructed isotope multiplicity is made from the choice of the input excitation energy for the shape of the LP multiplicity distribution calculation of GEMINI. For the errors shown in Fig.\ref{fig:Isotopic_distribution} the additional errors are evaluated from the maximum multiplicity difference between the calculations with the excitation energy between 2.0 and 2.5 MeV/nucleon.
%\textcolor{blue}{
In order to show the sensitivity of the selection of the GEMINI input excitation energy, all sigma values are artificially changed between $0.75\sigma$ to $1.25\sigma$ where $\sigma$ is calculated for $E_x=2.25 MeV/nucleon$. This is more or less the range of $\sigma$ values when $E_x$ is changed from 2.0 to 2.5 $MeV/nucleon$. The results are shown in Fig.~\ref{fig:Sigma}. As one can see, only minor changes of the multiplicity distribution are observed. One should note that in the actual simulations with different input excitation energy, the variation of $\sigma$ is more or less random and therefore the observed effect is smaller.
%}

%\textcolor{blue}{
As seen in Fig.\ref{fig:Isotopic_distribution_1}, the reconstructed hot isotopic distributions are quite well reproduced by those of the AMD simulations with g0 and g0AS interactions, whereas those of the g0ASS show a slightly wider distribution. It is interesting to note that the g0ASS results show better fit to the experimental secondary isotope distribution shown in Fig.\ref{fig:Secondary}. This better fits are "accidental" and caused by two factors, one from the higher excitation energy evaluation of the primary hot isotopes in the AMD simulations, as discussed in Ref.~\cite{Rodrigues13} and the other from the over prediction of the primary isotope distribution as seen above. In the AMD simulations, isotopes have the excitation energy in the range of $3-4 MeV/nucoen$. Whereas the evaluated experimental excitation energies are about 1 Mev/nucleon lower, depending on the isotopes. The wider distribution and the higher excitation energy, are more or less canceled out the yields of the cold isotopes and results in better fits for the g0ASS interaction in the secondary cold fragments. This fact indicates that it is important to separate the primary and secondary processes experimentally in order to refine the model simulations.
%}

\section*{IV. Power law distribution}

\begin{figure}[htb]
\includegraphics[scale=0.4]{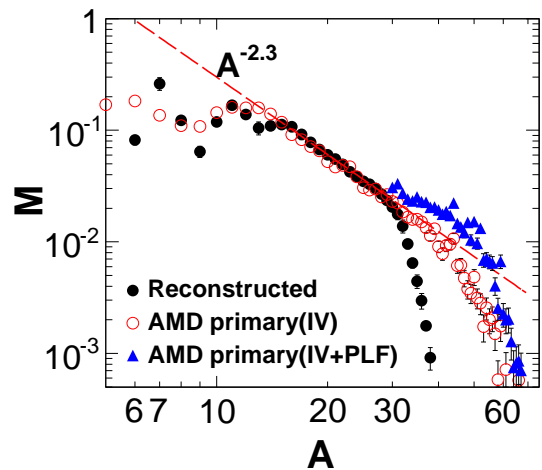}
\caption{\footnotesize
(Color online) The absolute multiplicities of the reconstructed hot isotopes are plotted by dots as a function of A together with those of the AMD primary isotopes of IV source(open circles). Those of the IV+PLF sources(solid triangles) are plotted only for A > 30, where one can see a clear deviation from those of the IV source.
}			
\label{fig:power_law}
\end{figure}

The multiplicity of the reconstructed primary isotopes are plotted as a function of A (dots) in Fig.\ref{fig:power_law}, together with those of the AMD primary isotopes obtained with the g0AS interaction (open circles). The multiplicities are given in absolute scale. The AMD multiplicitiesare the IV source component and calculated by minimizing the projectile-like and target-like components as mentioned earlier. The yields of isotopes with A $\geq$ 15 are well fitted by a power law distribution of $A^{-2.3}$ both for the reconstructed and AMD results. The fall off at A $>$ 30 in the reconstructed results is caused by the limitation of the available isotopes, which can be used for the reconstruction (Z $\leq$ 14). The associated LP multiplicities for Z $>$ 14 were not extracted in this work, because of their low yields.
The deviation from $A^{-2.3}$ in the AMD results at A $>$ 30 is partially caused by the selection of the IV source in the approximate  method. In the method most of the IV isotopes with $A > 30$ are gradually excluded by the angle selection condition $\theta_{lab} > 5^o$, because the heavier fragments are focused at forward angles as A becomes larger. For isotopes with $A > 30$, it is very difficult to isolate the IV component from the PLF one in the approximate method. In order to show the effect of the angular condition, the yields of the IV + PLF components ($\theta < 25^o$) are plotted by solid triangles for $A > 30$. The yields show the power law distribution with $A^{-2.3}$ roughly up to $A\sim55$, with a slight overestimation from the PLF contribution.

The power law result is consistent with the previous power law prediction in Ref.~\cite{Huang10_2}, though in that work the power law of $A^{-2.3}$ is predicted for all isotopes with A $\geq$ 1.
A significant deviation from the power law distribution of $A^{-2.3}$ is observed for the isotopes with A $<$ 15 in Fig. \ref{fig:power_law} both for the reconstructed hot and the AMD primary isotopes. The reason of the flattering of the mass distribution below A=15 is not clear at this moment.
The power law distribution observed in the AMD simulations should also be interpreted cautiously.
Furuta et al. demonstrated in Ref.~\cite{Furuta09} that in AMD calculations, IMFs are formed in a wide range of time interval (100 $fm/c$ - 300 $fm/c$) and the isotope yield distribution changes with time. However the yield and excitation energy distributions as a function of mass at a given time can be identified as one of statistically equilibrated ensembles generated by AMD separately. The temperature and density of the corresponding ensembles decrease monotonically in time.
In Ref.~\cite{Ono03}, they presented that isoscaling is hold in the the AMD events, which is not evident a priori for the dynamical models.
Their study, therefore, may indicate that the variety of the fragmentation process in AMD originate from the fluctuation of a statistical ensemble (a freeze-out ensemble) in time, density and temperature. This large fluctuation may cause difficulty in identifying a single freeze-out source and time on an event by event basis. The existence of such a freeze-out source is assumed in all statistical multifragmentation models and they can reproduce the experimental observables reasonably well as mentioned earlier. This fact and the observation of Refs.~\cite{Furuta09,Ono03} suggests that the multifragmentation in the AMD simulations reflects a large fluctuation of the virtual "freezeout" in space, density and time and causes a variety of cluster generation at early stages of the reaction. The experimental observation of the power law distribution for A $\geq$ 15 may suggest that there ia a virtual "freezeout" volume for the production of the heavier fragments, but for the production of the lighter fragments dynamical processes, such as semi-transparency~\cite{Wada00,Wada04}, neck-emissions and so on, become more important.

\section*{V. Characteristic properties of the fragmenting source}

The characteristic properties of the fragmenting source have been studied through the production of IMFs, using the Modified Fisher Model (MFM)~\cite{Fisher1967,Minich1982,Hirsch1984,Bonasera2008}.
MFM is applied to characterize the emitting source of IMFs in the previous works~\cite{Bonasera2008,Huang10_1,Huang10_2,Huang10_3,Lin14,Liu14}.
In the framework of MFM, the yield of an isotope with $I=N-Z$ and $A$ (N neutrons and Z protons) produced in a multi-fragmentation reaction, can be given as
\begin{equation}
\begin{split}
Y(I,A) =& Y_{0} \cdot A^{-\tau}exp[\frac{W(I,A)+\mu_{n}N+\mu_{p}Z}{T}\\
&+S_{mix}(I,A)].
\end{split}
\label{eq:eq_MFM}
\end{equation}
Using the generalized Weizs$\ddot{a}$cker-Bethe semiclassical mass formula~\cite{Weizsacker1935,Bethe1936}, $W(N,Z)$ can be approximated as
\begin{equation}
\begin{split}
W(I,A) =& a_{v}A- a_{s}A^{2/3}- a_{c}\frac{Z(Z-1)}{A^{1/3}}\\
&-a_{sym}\frac{I^{2}}{A}
- a_{p}\frac{\delta}{A^{1/2}},\\
\delta =& - \frac{(-1)^{Z}+(-1)^{N}}{2}.
\end{split}
\label{eq:eq_WB}
\end{equation}
In Eq.\eqref{eq:eq_MFM}, $A^{-\tau}$ and $S_{mix}(I,A)=Nln(N/A)+Zln(Z/A)$ originate from the increases of the entropy and the mixing entropy at the time of the fragment formation, respectively. $\mu_{n}$ ( $\mu_{p}$ ) is the neutron (proton) chemical potential. $\tau$ is the critical exponent. In this work, the value of $\tau=2.3$ is adopted from the previous studies~\cite{Bonasera2008}.
In general coefficients, $a_{v}$, $a_{s}$, $a_{sym}$, $a_{p}$ and the chemical potentials are temperature and density dependent, even though these dependence are not shown explicitly.

When one makes a yield ratio between isobars from Eq.\eqref{eq:eq_MFM}, A dependent parts are canceled out. Especially when the isobars differing 2 unit in I are used, one can get the following equation.
%\begin{widetext}
\begin{eqnarray}
R(I+2,I,A) = Y(I+2,A)/Y(I,A)   \nonumber\\
=  exp\{[\mu_{n}- \mu_{p}+ 2a_{c}(Z-1)/A^{1/3}-\nonumber\\ 4a_{sym}(I+1)/A-\delta(N+1,Z-1) \nonumber\\
- \delta(N,Z)]/T+ \Delta(I+2,I,A)\},
\label{eq:eq_RI2}
\end{eqnarray}
%\end{widetext}
where $\Delta(I+2,I,A)=S_{mix}(I+2,A) - S_{mix}(I,A)$.
When the above equation is applied to the isobars with $I = N - Z = -1$ and 1, then the symmetry energy term and pairing term drop out and the following equation is obtained.
\begin{equation}
    \ln[R(1, -1, A)]=[\Delta\mu+2a_{c}(Z-1)/A^{1/3}]/T
\label{eq:Coulomb}
\end{equation}
where $\Delta\mu=(\mu_{n}-\mu_{p})$.

\begin{figure}[htb]
\includegraphics[scale=0.4]{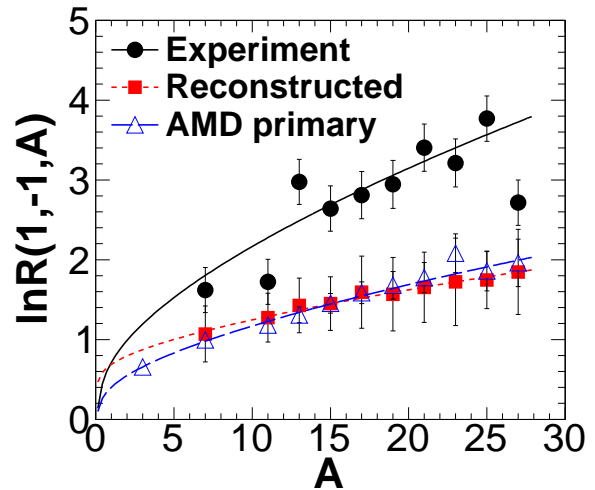}
\caption{\footnotesize
(Color online) $\ln[R(1, -1, A)]$ values are plotted as a function of fragment mass number A for the experimental (dots) and reconstructed hot fragments (closed squares) as well as the primary fragments of the AMD simulation (open triangles). The curves are obtained by free parameter search of  $\Delta\mu/T$ and $a_{c}/T$ in Eq.(\ref{eq:Coulomb}). The extracted parameter from the reconstructed data is $a_{c}/T=0.14\pm0.04$, using $\Delta\mu/T=0.67$. The values used for the experimental and AMD primary data are from Ref.~\cite{Huang10_1} and the values used are $(\Delta \mu/T,a_{c}/T)=(0.71,0.35)$ for the experiment, and $(0.40,0.18)$ for the AMD primary.
}			
\label{fig:Coulomb_parameter}
\end{figure}

$ln{R(1,-1,A)}$ values in Eq.(\ref{eq:Coulomb}) are shown as a function of A in Fig.\ref{fig:Coulomb_parameter} for the experimental cold isotopes, those from the reconstructed hot ones extracted in the previous section and those from the AMD primary ones. Following the procedure described in Ref.~\cite{Huang10_1} $a_{c}/T$ = 0.35 from the experimental cold isotopes and $a_{c}/T$ = 0.18 from the primary fragments of the AMD simulation with g0AS were obtained and the fits curves are shown by solid and dashed lines in the figure, respectively. For the reconstructed hot isotopes, using $a_{c}/T$ and $\Delta \mu/T$ as free parameters, $a_{c}/T=0.14 \pm 0.04$ and $\Delta \mu/T$ = 0.67 are obtained and shown by the dotted line.  The results from the reconstructed data show significant difference from those of the experimental cold multiplicities and distribute close to those of the AMD primary multiplicities, which is an indication of the sequential decay effect on the Coulomb parameter in Eq.(\ref{eq:Coulomb}).

In order to further study the characteristic properties of the  source of the primary isotopes, the ratio of the symmetry energy coefficient relative to the temperature, $a_{sym}/T$, is examined. In a similar way to that of  Eq.(\ref{eq:Coulomb}),
the $a_{sym}/T$ value can be extracted using the yield ratio of three isobars with $I=-1, 1$ and 3 as,
\begin{eqnarray}
a_{sym}/T &= -\frac{A}{8}\{\ln[R(3, 1, A)]-\ln[R(1, -1, A)] \nonumber\\
&-\Delta(3, 1, A) + \Delta E_{c}\}
\label{eq:Symmetry}
%\end{equation}
\end{eqnarray}
$\Delta(3, 1, A)$ is the difference in mixing entropies of isobars A with $I = 3$ and 1. $\Delta E_{c}$ is the difference of the Coulomb energy between the neighboring isobars and given by $\Delta E_{c} = 2a_{c}/(A^{1/3}T)$. The $a_{c}$ value is obtained from the above analysis used in Fig.\ref{fig:Coulomb_parameter}. One should note that the values of $\Delta(3, 1, A)$ and $\Delta E_{c}$ are small compared to the first two terms and they have opposite signs each other.

In a transport model such as AMD, the dynamic evolution of the system is such that variations in the temperature, density and symmetry energy are closely correlated with each other. If one of these parameters is determined, then other parameters can be extracted in a self-consistent manner from the transport model solutions using these relationships. In the following the experimentally extracted $a_{sym}/T$ values from the reconstructed isotopes, are compared with those from the AMD simulations using
g0, g0AS and g0ASS interactions. From the comparisons,
the density of the fragmenting source is determined and then the temperature and symmetry energy are extracted using the model predicted correlations. This method has been applied in Refs.~\cite{Lin14,Liu14}.

\begin{figure}[htb]
\includegraphics[scale=0.4]{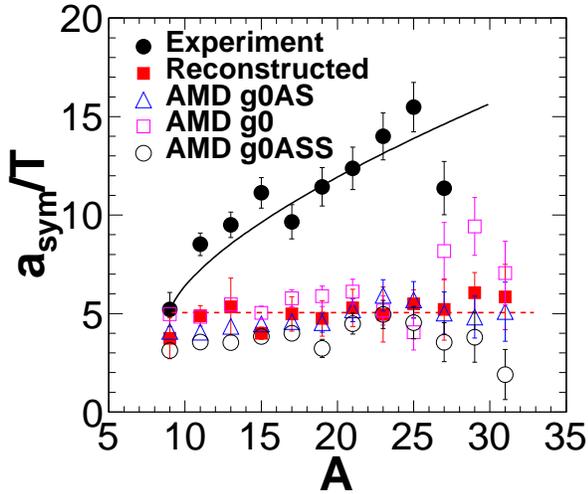}
\caption{\footnotesize
(Color online) The calculated values of $a_{sym}/T$, are plotted as a function of $A$ from the experimental cold isotopes (dots) and the reconstructed hot isotopes (solid squares). The solid curve is from Ref.~\cite{Huang10_1}. The dashed curve is the average value of those from the reconstructed ones. AMD results are shown by open symbols, those for g0 (squares), for g0AS (triangles) and for g0ASS (circles). The average $a_{sym}/T$ values for those from the reconstructed and the AMD simulations are given in the fourth column of Table~\ref{table:data_results}.
}			
\label{fig:Symmetry_parameter}
\end{figure}

Using Eq.(\ref{eq:Symmetry}), $a_{sym}/T$ values were calculated and the results are shown in Fig.\ref{fig:Symmetry_parameter}.
The results from the reconstructed primary isotopes(solid squares) show a rather flat distribution and a significant difference from those for the experimentally observed cold isotopes (dots), indicating that the strong mass dependence of the latter originates from the secondary cooling process as concluded in Ref.~\cite{Huang10_1}.
AMD results with the three different interactions show a similar flat distribution to those of the reconstructed ones. Their distributions are more or less parallel to each other, but have different values. Their average values are given in the fourth column of Table~\ref{table:data_results}.

\begin{figure}[htb]
\includegraphics[scale=0.4]{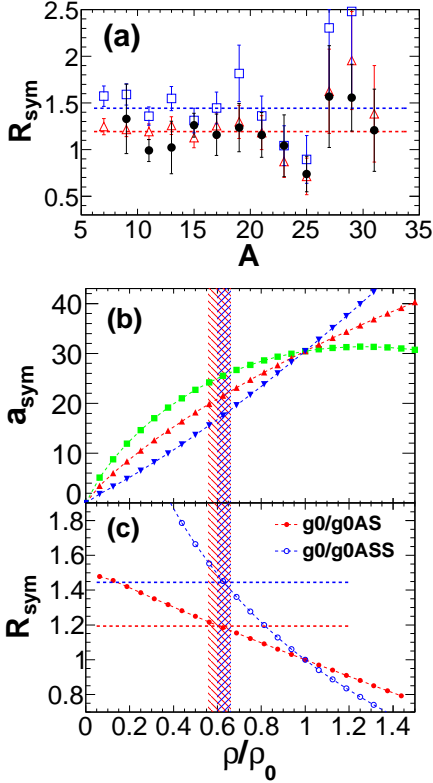}
\caption{\footnotesize
(Color online) (a) Ratios of the calculated $a_{sym}/T$ values for g0/g0AS (open triangles), g0/g0ASS (open squares) and g0/the reconstructed experimental yield (dots). Dotted lines are the average values for the AMD simulations. The values are given in the first column of Table~\ref{table:data_results}. (b) Symmetry energy coefficient vs density for g0, g0AS and g0ASS used in the AMD simulations. The shaded vertical area indicates the density of fragmenting system extracted from the ratio of the symmetry energy coefficient. Two different shade patterns are used for the density values given in the second column of Tabel~\ref{table:data_results}. (c) Ratio of the symmetry energy coefficients, used in (b), between /g0g0AS and g0/g0ASS as a function of the density. The horizontal dotted lines indicates the ratio values extracted from the $a_{sym}/T$ values in (a).
}			
\label{fig:Symmetry_AMD}
\end{figure}

The ratios $a_{sym}/T$ for g0 relative to those for g0AS and g0ASS are plotted in Fig.\ref{fig:Symmetry_AMD}(a), together with the ratio of those from g0 relative to those from the reconstructed yields (dots). Both of the calculated ratios are more or less constant as a function of $A$, though those from the reconstructed yields have a slightly larger fluctuation than those of the simulations. The average values of these ratios are given in the first column of Table~\ref{table:data_results}.
Following Ref.~\cite{Ono03}, we interpret the ratios as resulting from the difference of the symmetry energy coefficient at the density and temperature of fragment formation. In Fig.\ref{fig:Symmetry_AMD}(b), the density dependence of the symmetry energy coefficient for g0, g0AS and g0ASS is shown as a function of $\rho / \rho_{0}$. In Fig.\ref{fig:Symmetry_AMD}(c), their ratios for g0/g0AS and g0/g0ASS are shown.
From the ratio values of the simulations in Fig.\ref{fig:Symmetry_AMD}(a), the corresponding densities are extracted as indicated by the vertical shade areas in Fig.\ref{fig:Symmetry_AMD}(b) and (c). The extracted values are $\rho / \rho_{0}= 0.61\pm 0.05$ for g0/g0AS and $0.63\pm 0.03$ for g0/g0ASS. These are given in the second column of Table~\ref{table:data_results}. The error becomes smaller for g0/g0ASS because the ratio of g0/g0ASS shows a sharper slope as a function of the density and therefore greater sensitivity to the density dependence. Assuming the nucleon density is same for the three different interactions used, the nucleon density of the fragmenting system is determined from the overlap value of the extracted values as $\rho / \rho_{0} = 0.63 \pm 0.03$. This assumption is reasonable because the nucleon density is mainly determined by the stiffness of the EOS and not by the density dependence of the symmetry energy term.

The corresponding symmetry energy coefficient values from the calculations are extracted from Fig.\ref{fig:Symmetry_AMD}(b) as $25.7\pm0.6$, $21.2\pm1.2$, and $ 17.8\pm0.9$ MeV for g0, g0AS and g0ASS, respectively. These values are given in the third column of Table~\ref{table:data_results}.
For the AMD simulations, the temperature, $T=a_{sym}/(a_{sym}/T)$, is calculated. We find $T=4.9\pm 0.2 , 4.9\pm 0.4$ and $5.1\pm 0.5 MeV$ respectively for the g0, g0AS and g0ASS interactions. The temperatures appear in the fifth column of Table~\ref{table:data_results}.
One should note that the errors for the temperature and symmetry energy values originate from those on the density values and the $a_{sym}/T$ values in the first and fourth columns of Table~\ref{table:data_results}, since they are determined using their predicted correlations in the AMD model.

From the temperature values for the AMD simulations with different interactions, the temperature   for the fragmenting source is determined from  the overlap  values, assuming the same source temperature for the different density dependencies of the symmetry energy coefficient. The overlap value is $T = 4.9 \pm 0.2 MeV$. Using this temperature and the experimental $a_{sym}/T$ value in the bottom of the fourth column, the experimental symmetry energy coefficient is determined as $a_{sym} = 24.7 \pm 1.9 MeV$.
The extracted symmetry energy coefficient, temperature and density for the fragment formation show notable differences from those of Ref.~\cite{Shetty07}, where the values were extracted from the experimentally observed secondary yields using isoscaling parameters. In that work, the reactions of $^{40}$Ar, $^{40}$Ca + $^{58}$Ni,$^{58}$Fe at 25-55 MeV/nucleon were studied. Isoscaling parameters were extracted from the experiments and compared to those of the AMD and SMM simulations using interactions with different density dependencies of the symmetry energy. From those comparisons, the values listed in the bottom three rows of Table~\ref{table:data_results} were obtained.

\begin{table}[ht]
\caption{Extracted parameters. The lines indicated with $^{*}$ at the bottom three rows are from Ref.~\cite{Shetty07} and E$_{x}$ is given in MeV. } % title of Table
\centering % used for centering table
\begin{tabular}{c c c c c c} % centered columns (6 columns)
\hline\hline %inserts double horizontal lines
 & Ratio & $\rho/\rho_{0}$ & $a_{sym}$ & $a_{sym}/T$ & T\\ [0.5ex] % inserts table
 &       &                 & (MeV)     &             & (MeV)\\ [0.5ex] % inserts table
%heading
\hline % inserts single horizontal line
g0    &               &               & 25.7$\pm$0.6 & 5.29$\pm$0.13 & 4.9$\pm$0.2\\
g0/g0AS& 1.19$\pm$0.03 & 0.61$\pm$0.05 &  &  & \\
g0AS  &               &               & 21.2$\pm$1.2 & 4.31$\pm$0.12 & 4.9$\pm$0.4\\
g0/g0ASS& 1.44$\pm$0.05 & 0.63$\pm$0.03 &  &  & \\
g0ASS &               &               & 17.8$\pm$0.9 & 3.50$\pm$0.12 & 5.1$\pm$0.5\\
\hline %inserts single line
Exp   &               & 0.63$\pm$0.03 & 24.7$\pm$1.9 &  5.04$\pm$0.32  & 4.9$\pm$0.2\\[0.5ex]
\hline\hline %inserts double horizontal lines
%Ref.~\cite{Shetty07} &    & 0.45$\pm$0.12 & 17$\pm$2  &              & 6.5$\pm$0.5\\[0.5ex]
E$_{x}$=5$^{*}$     &    & 0.50$\pm$0.12 & 20$\pm$2  &              & 5.7$\pm$0.5\\[0.5ex]
E$_{x}$=7.5$^{*}$  &    & 0.45$\pm$0.12 & 17$\pm$2  &              & 6.5$\pm$0.5\\[0.5ex]
E$_{x}$=9.5$^{*}$   &    & 0.30$\pm$0.12 & 16$\pm$2  &              & 7.0$\pm$0.5\\[0.5ex]
\hline %inserts single line
\end{tabular}
\label{table:data_results} % is used to refer this table in the text
\end{table}

\section*{VI. SUMMARY}

The multiplicity distribution of primary hot isotopes was experimentally reconstructed for fragments produced in the  $^{64}$Zn + $^{112}$Sn reaction at 40 MeV/nucleon. A kinematical focussing technique was employed to isolate particles emitted from the primary fragments. Using the experimental multiplicities of isotopically identified detected fragments and their associated LP multiplicities together with LP distributions widths from a GEMINI simulation, a Monte Carlo method was used for the reconstruction.
The multiplicity distributions of the reconstructed primary fragments  are in good agreement with those calculated from the AMD with g0 or g0AS interactions. The results for g0ASS exhibit a slightly wider distribution in neutron number.
The mass yields of the reconstructed hot isotopes for $A \ge 15$ show a power law distribution of $A^{-2.3}$, whereas those with $A < 15$ show a significant deviation from that, suggesting that the production mechanism for these lighter isotopes are different from those of the heavier ones. This power law behavior together with other statistical natures may reflect the fact that there is a virtual "freezeout" in transport models and a large fluctuation in space and time causes a variety of cluster generation at early stages of the reaction.

%In order to study the phase transition in the fragmenting source, the Landau free energy approach is applied. The fit result  for the experimentally  detected isotopes show significant differences from that for the primary fragments, indicating that the Landau free energies are significantly affected by the secondary decay process. No conclusive results are obtained from the reconstructed primary isotope yields for the first-order phase transition.

The ratios of the symmetry energy coefficients to the temperature, $a_{sym}/T$, extracted based on MFM, were utilized to determine the density, temperature and symmetry energy coefficient at the time of the fragment formation in a self-consistent way. From the comparisons with AMD simulations using different interactions, , $\rho / \rho_{0} = 0.63 \pm 0.03 $, a temperature of $T = 4.9 \pm 0.2 MeV$ and the symmetry energy coefficient of $a_{sym} = 24.7 \pm 1.9 MeV$ are extracted at the time of the reconstructed primary isotope formation.

\section*{Acknowledgments}

We thank the staff of the Texas A$\&$M Cyclotron facility for their support during the experiment. We thank the Institute of Nuclear Physics of the University of Louvain and Prof. Y. El Masri for allowing us to use the DEMON detectors. We thank A. Ono and R. J. Charity for providing their codes. This work is supported by the U.S. Department of Energy under Grant No. DE-FG03-93ER40773 and the Robert A. Welch Foundation under Grant A0330. This work is also supported by the National Natural Science Foundation of China (Grants No. 11075189 and No. 11105187)(I ADD FUFEN'S FOUNDATION) and 100 Persons Project (Grants No. 0910020BR0 and No. Y010110BR0), ADS project 302 (Grants No. Y103010ADS) of the Chinese Academy of Sciences. One of the author (R.W) thanks the program of the visiting professorship of senior international scientists of the Chinese Academy of Science" for their support.

%\end{acknowledgments}
%%\label{}



%% References


%\bibliographystyle{elsarticle-num}

\begin{thebibliography}{00}

+
\bibitem{Bohr36} N. Bohr, Nature {\bf 137}, 35 (1936).
\bibitem{Borderie08} B. Borderie and M. F. Rivet, Prog. Part. Nucl. Phys. {\bf 61}, 551 (2008).
\bibitem{Gulminelli06} F. Gulminelli {\it et al.},Eur.Phys.J.A30,1(2006) and related topics in the volume.
\bibitem{Chomaz04} Ph. Chomaz {\it et al.} Phys. Rep. 389, 263 (2004).
\bibitem{Finn82} J. E. Finn et al., Phys. Rev. Lett. 49, 1321 (1982).
\bibitem{Minich1982}R. W. Minich {\it et al.}, Phys. Lett. {\bf B118}, 458 (1982).
\bibitem{Hirsch1984}A. S. Hirsch {\it et al.}, Nucl. Phys. {\bf A418}, 267c (1984).
%\bibitem{Chomaz06} P. Chomaz et al., Eur. Phys. J. A 30, 1 (2006).
\bibitem{Elliott00} J. B. Elliott et al., Phys. Rev. C 62, 064603 (2000); J. B. Elliott et al., Phys. Rev. Lett. 88, 042701 (2002); J. B. Elliott et al., Phys. Rev. C 67, 024609 (2003).
\bibitem{Lattimer04} J. M. Lattimer and M. Prakash, Science {\bf 23},536 (2004).
\bibitem{BALi08} B. A. Li {\it et. al.} Phys.Rep. {\bf 464},113 (2008).
\bibitem{Tsang2001}M. B. Tsang {\it et al.}, Phys. Rev. Lett. {\bf 86}, 5023 (2001).
\bibitem{Tsang2004}M. B. Tsang {\it et al.}, Phys. Rev. Lett. {\bf 92}, 062701 (2004).
\bibitem{Famiano2006}M. A. Famiano {\it et al.}, Phys. Rev. Lett. {\bf 97}, 052701 (2006).
\bibitem{Li2007}T. Li {\it et al.}, Phys. Rev. Lett. {\bf 99}, 162503 (2007).
\bibitem{Klimkiewicz2007}A. Klimkiewicz {\it et al.}, Phys. Rev. {\bf C76}, 051603 (2007).
\bibitem{Trippa2008}L. Trippa {\it et al.}, Phys. Rev. {\bf C77}, 061304 (2008).
\bibitem{zak2012}Z. Kohley {\it et al.}, Phys. Rev. {\bf C85}, 064605 (2012).
\bibitem{Xu2000}H. S. Xu {\it et al.}, Phys. Rev. Lett. {\bf 85}, 716 (2000).
\bibitem{Tsang2001_1}M. B. Tsang {\it et al.}, Phys. Rev. {\bf C64}, 054615 (2001).
\bibitem{Huang2010}M. Huang {\it et al.}, Nucl. Phys. {\bf A847}, 233 (2010).
\bibitem{Aichelin85}J. Aichelin {\it et al.}, Phys. Rev. {\bf C31}, 1730 (1985).
\bibitem{Colonna98} M. Colonna {\it et al.}, Nucl. Phys. {\bf A642}, 449 (1998).
\bibitem{Baran12} V. Baran, M. Colonna, M. Di Toro and R.Zus, Phys. Rev. {\bf C85}, 054611 (2012).
\bibitem{Gagnon12} F. Gagnon-Moisan et al., Phys. Rev. {\bf C86}, 044617 (2012).
\bibitem{Kruse85} H. Kruse {\it et al.}, Phys. Rev. {\bf C31}, 1770 (1985).
\bibitem{Baran02}V. Baran V {\it et a;.},Nucl. Phys. A703, 603 (2002).
\bibitem{Peilert89} G. Peilert {\it et al.}, Phys. Rev. {\bf C39}, 1402 (1989).
\bibitem{Aichelin91} J. Aichelin, Phys. Rep. {\bf 202}, 233 (1991).
\bibitem{Lukasik93} J. Lukasik {\it et al.}, Acta Phys. Polon. {\bf B24}, 1959 (1993).
\bibitem{Papa01} M. Papa, T. Maruyama, A. Bonasera, Phys. Rev. {\bf C64}, 024612 (2001).
\bibitem{Papa05} M. Papa, G. Giuliani, A. Bonasera, J. Comput. Phys. {\bf 208}, 403 (2005).
\bibitem{Papa07} M. Papa {\it et al.}, Phys. Rev. {\bf C75}, 054616 (2007).
\bibitem{Papa09} M. Papa, G. Giuliani, Eur. Phys. J. {\bf A39}, 117 (2009).
\bibitem{Wang02} N. Wang, Z.Li and X.Wu Phys. Rev. {\bf C65} 064608 (2002).
\bibitem{Wang04} N. Wang {\it et al.}, Phys. Rev. {\bf C69}, 034608 (2004).
\bibitem{Zhang05} Y. Zhang and Z. Li Phys. Rev. {\bf C71}, 024604 (2005).
\bibitem{Zhang06} Y. Zhang and Z. Li, Phys. Rev. {\bf C74}, 014602 (2006).
\bibitem{Zhang12} Y. Zhang {\it et al.}, Phys. Rev. {\bf C85}, 024602 (2012).
\bibitem{Feldmeier90} H. Feldmeier, Nuclear Phys. {\bf A515}, 147 (1990).
\bibitem{Ono96} A. Ono and H. Horiuchi, Phys. Rev. {\bf C53}, 2958 (1996).
\bibitem{Ono99} A. Ono, Phys. Rev. {\bf C59}, 853 (1999).
\bibitem{Ono02} A. Ono, S. Hudan, A. Chbihi, J. D. Frankland, Phys. Rev. {\bf C66}, 014603 (2002).
%\bibitem{Gregoire87} C. Gregoire {\it et al.}, Nuclear Phys. {\bf A465}, 317 (1987).
%\bibitem{Bertsch88} G. F. Bertsch {\it et al.}, Phys. Rep. {\bf 160}, 189 (1988).
%\bibitem{Bonasera94} A. Bonasera {\it et al.}, Phys. Rep. {\bf 243}, 1 (1994).
\bibitem{Gross90} D. H. E. Gross, Rep. Prog. Phys. {\bf 53}, 605 (1990).
\bibitem{DAgostino99} M. D'Agostino {\it et al.}, Nucl. Phys. {\bf A652}, 359 (1999).
\bibitem{Bondorf85} J. P. Bondorf, R. Donangelo, I. N. Mishustin, Nucl. Phys. {\bf A443}, 321 (1985).
\bibitem{Bondorf95} J. Bondorf, A. S. Botvina, A. S. Iljinov, I. N. Mishutin, K. Sneppen, Phys. Rep. {\bf 257}, 133 (1995).
\bibitem{Botvina95} A. S. Botvina {\it et al.}, Nucl. Phys. {\bf A584}, 737 (1995).
\bibitem{DAgostino96} M. D'Agostino {\it et al.}, Phys. Lett. {\bf B371}, 175 (1996).
\bibitem{Scharenberg01} R. P. Scharenberg {\it et al.}, Phys. Rev. {\bf C64}, 054602 (2001).
\bibitem{Bellaize02} N. Bellaize {\it et al.}, Nucl. Phys. {\bf A709}, 367 (2002).
\bibitem{Avdeyev02} S. P. Avdeyev {\it et al.}, Nucl. Phys. {\bf A709}, 392 (2002).
\bibitem{Ogul12} R. Ogul {\it et al.}, Pyhs. Rev. {\bf C83}, 024608 (2011).
%\bibitem{Rizzo07} J. Rizzo {\it et al.}, Phys. Rev. {\bf C76}, 024611 (2007).
%\bibitem{Colonna10} M. Colonna {\it et al.}, Phys. Rev. {\bf C82}, 054613 (2010).
\bibitem{Marie98}N. Marie {\it et al.}, Phys. Rev. C {\bf 58}, 256 (1998).
\bibitem{Hudan03}S. Hudan {\it et al.}, Phys. Rev. C {\bf 67}, 064613 (2003)
\bibitem{Rodrigues13} M. R. D. Rodrigues {\it et al.}, Phys. Rev. {\bf C88}, 034605 (2013).
\bibitem{Lin14} W. Lin {\it et al.}, Phys. Rev. {\bf C89}, 021601(R) (2014).
\bibitem{Huang10_1} M. Huang {\it et al.}, Phys. Rev. {\bf C81}, 044620 (2010).
\bibitem{Huang10_2} M. Huang {\it et al.}, Phys. Rev. {\bf C82}, 054602 (2010).
\bibitem{Chen10} Z. Chen {\it et al.}, Phys. Rev. {\bf C81}, 064613 (2010).
\bibitem{Hubert90} F. Hubert, R. Bimbot, and H. Gauvin, At. Data Nucl. Data Tables {\bf 46}, 1, (1990).
\bibitem{Tilquin95} I. Tilquin {\it et al.}, Nucl. Instr. and Meth. {\bf A365}, 446 (1995).
\bibitem{Zhang13}S. Zhang {\it et al.}, Nucl. Instr. and Meth. A. {\bf 709}, 68 (2013)
\bibitem{Charity88} R. J. Charity {\it et al.}, Nucl. Phys. {\bf A483}, 371 (1988).
\bibitem{Wada04}R. Wada {\it et al.}, Phys. Rev. {\bf C69}, 044610 (2004).
\bibitem{Awes81} T. C. Awes {\it et al.}, Phys. Rev. {\bf C24}, 89 (1981).
\bibitem{Ono03} A. Ono, P. Danielewicz, W. A. Friedman, W. G. Lynch, and M. B. Tsang, Phys. Rev. {\bf C68},  051601(R) (2003). For g0AS, x=-1/2 and for g0ASS x=-2 are used in Eq.(2) in the reference.
\bibitem{Furuta09} T. Furuta and A. Ono, Phys. Rev. {\bf C79}, 014608 (2009).
\bibitem{Wada00} R. Wada {\it et al.}, Phys. Rev. {\bf C62}, 034601 (2000).
%\bibitem{Huang87}K. Huang, Statistical Mechanics (Wiley \& Sons, New York, 1987), 2nd edition, Chap. 16, 17.
\bibitem{Huang10_3} M. Huang {\it et al.}, Phys. Rev. {\bf C81}, 044618 (2010).
%\bibitem{Tripathi11} R. Tripathi {\it et al.}, Phys. Rev. {\bf C83}, 054609 (2011).
%\bibitem{Mabiala13} J. Mabiala {\it et al.}, Phys. Rev. {\bf C 87}, 017603 (2013).
\bibitem{Bonasera2008}A. Bonasera {\it et al.}, Phys. Rev. Lett. {\bf 101}, 122702 (2008).
\bibitem{Fisher1967}M. E. Fisher, Rep. Prog. Phys. {\bf 30}, 615 (1967).
\bibitem{Liu14} X. Liu {\it et al.}, Phys. Rev. {\bf C90}, 014605 (2014).
\bibitem{Weizsacker1935}C. F. von Weizs$\ddot{a}$cker, Z. Phys. {\bf 96}, 431 (1935).
\bibitem{Bethe1936}H. A. Bethe, Rev. mod. Phys. {\bf 8}, 82 (1936).

%\bibitem{Natowitz02} J. B. Natowitz, R. Wada, K. Hagel, T. Keutgen, M. Murray, A. Makeev, L. Qin, P. Smith, and C. Hamilton, Phys. Rev. {\bf C65}, 034618 (2002).
%\bibitem{Weizsacker35}C. F. V. Weizs\"acker, Z. Phys. 96, 431 (1935).
%\bibitem{Jiang12}  H. Jiang, G. J. Fu, Y. M. Zhao and A. Arima, Phys. Rev. {\bf C85}, 024301 (2012)
%\bibitem{Ono04} A. Ono, P. Danielewicz, W. A. Friedman, W. G. Lynch, and M. B. Tsang, Phys. Rev. {\bf C70},  041604(R) (2004)
\bibitem{Shetty07} D. V. Shetty, S. J. Yennello, and G. A. Souliotis, Phys. Rev. {\bf C76}, 024606 (2007).

\end{thebibliography}



\end{document}